\documentclass[pmlr,twocolumn,10pt]{jmlr}
\pdfoutput=1
\usepackage[english]{babel}
\usepackage[utf8]{inputenc}
\usepackage{multicol,multirow,array}
\usepackage{amsmath,xspace,booktabs,float,tocloft,mathtools}

\newtheorem{goal}{Goal}[section]
\DeclarePairedDelimiter\paren{\lparen}{\rparen}

\usepackage{graphicx}
\graphicspath{{figures/}{.}}
\usepackage[hyphenbreaks]{breakurl}

\usepackage[draft,inline,nomargin,index]{fixme}
\fxsetup{theme=colorsig,mode=multiuser,inlineface=\itshape,envface=\itshape}
\FXRegisterAuthor{sv}{asv}{Suresh}
\FXRegisterAuthor{de}{ade}{Dani}
\FXRegisterAuthor{sf}{asf}{Sorelle}
\FXRegisterAuthor{cs}{ces}{Carlos}

\newcommand{\one}{A\xspace}
\newcommand{\two}{B\xspace}

\newcommand{\polya}{P\'{o}lya\xspace}

\newcommand{\myfrac}[2]{#1(#2)^{-1}}
\newcommand{\predpol}{\texttt{PredPol}\xspace}

\theorembodyfont{\upshape}
\theoremheaderfont{\scshape}
\newtheorem{assume}{Assumption}[section]

\jmlrvolume{81}
\jmlryear{2018}
\jmlrworkshop{Conference on Fairness, Accountability, and Transparency}

\title{Runaway Feedback Loops in Predictive
  Policing\titletag{\thanks{This research was funded in part by the NSF under
      grants IIS-1633387, IIS-1513651, and IIS-1633724. Code for our urn simulations can be found at \url{https://github.com/algofairness/runaway-feedback-loops-src}.}}}

\author{\Name{Danielle Ensign} \Email{daniphye@gmail.com} \\
\addr University of Utah
\AND
\Name{Sorelle A. Friedler} \Email{sorelle@cs.haverford.edu} \\
\addr Haverford College
\AND
\Name{Scott Neville} \Email{drop.scott.n@gmail.com} \\
\addr University of Utah
\AND
\Name{Carlos Scheidegger} \Email{cscheid@cscheid.net}\\
\addr University of Arizona
\AND
\Name{Suresh Venkatasubramanian}\thanks{Corresponding author.} \Email{suresh@cs.utah.edu} \\
\addr University of Utah
}

\editors{Sorelle A. Friedler and Christo Wilson}

\begin{document}

\maketitle
\begin{abstract}
Predictive policing systems are increasingly used to determine
how to allocate police across a city in order to best prevent
crime. Discovered crime data (e.g., arrest counts) are used to help update
the model, and the process is repeated. Such systems have
been empirically shown to be susceptible to runaway feedback loops, where
police are repeatedly sent back to the same neighborhoods
regardless of the true crime rate.

In response, we develop a
mathematical model of predictive policing that proves why this feedback
loop occurs, show empirically that this model exhibits such
problems, and demonstrate how to change the inputs to a
predictive policing system (in a black-box manner) so the
runaway feedback loop does not occur, allowing the true crime
rate to be learned. Our results are quantitative: we can establish a link (in
our model) between the degree to which runaway feedback causes problems and the
disparity in crime rates between areas. Moreover, we can also demonstrate the
way in which \emph{reported} incidents of crime (those reported by residents) and \emph{discovered} incidents
of crime (i.e. those directly observed by police officers dispatched as a result
of the predictive policing algorithm) interact: in brief, while reported
incidents can attenuate the degree of runaway feedback, they cannot entirely
remove it without the interventions we suggest. 

\end{abstract}

\section{Introduction}

Machine learning models are increasingly being used to make real-world
decisions, such as who to hire, who should receive a loan, where to send police,
and who should receive parole.  These deployed models mostly use traditional
batch-mode machine learning, where decisions are made and observed results
supplement the training data for the next batch. 

However, the problem of \emph{feedback} makes  traditional batch learning frameworks both inappropriate and
(as we shall see) incorrect. Hiring algorithms only receive feedback on people who were hired, predictive
policing algorithms only observe crime in neighborhoods they patrol, and so
on. Decisions made by the system influence the data that is fed to it in the future. For
example, once a decision has been made to patrol a certain neighborhood, crime
discovered in \emph{that} neighborhood will be fed into the training apparatus for the
next round of decision-making. 



In this paper, we focus on predictive policing -- an important exemplar problem demonstrating these feedback issues.  Predictive policing is increasingly employed to determine where to send police, who to target for surveillance, and even 
who may be a future crime victim~\citep{RAND}.  We focus on the most popular of these forms of predictive policing (with \predpol, HunchLab, IBM, and other companies entering the market) which attempts to determine how to deploy police given historical crime data.

 \begin{definition}[Predictive Policing]
   Given historical crime incident data for a collection of regions, decide how to
   allocate patrol officers to areas to detect crime. 
 \end{definition}

Once police are deployed based on these predictions, data from observations in the neighborhood is then used to further update the model.  We will call these observations \emph{discovered incidents}, as opposed to \emph{reported incidents} that are crime incidents reported to the police (e.g., via 911 calls).  Since such discovered incidents only occur in neighborhoods that police have been sent to \emph{by the predictive policing algorithm itself}, there is the potential for this sampling bias to be compounded, causing a runaway feedback loop.  Indeed, \citet{LumIsaac2016} have shown that this can happen. 

Lum and Isaac's work focused on \predpol \citep{predpol}, a predictive policing
system in use by the LAPD and other cities across the U.S..  
\citet{LumIsaac2016} model what would happen if \predpol were used in Oakland to
distribute police to find drug crime by using historical crime incident data as the historical data and a synthetic population of likely drug users based on public
health data \cite{icpsr_ncvs, icpsr_nsduh}; they find that increasing policing efforts
based on discovered incidents causes \predpol's prediction to substantially diverge from the true crime rate, repeatedly sending back police to the same neighborhoods.

In addition to its importance in the criminal justice pipeline, predictive policing serves as an archetypal problem,
through which we can better understand issues which arise out of deploying batch-mode machine learning algorithms
in an online setting, where they essentially see results that are influenced by their own predictions.
Other such algorithms include recidivism prediction, hiring algorithms, college admissions,
and distribution of loans.  In all of these contexts, the outcome of the prediction (e.g., who to hire) determines
what feedback the algorithm receives (e.g., who performs well on the job).

\subsection{Results}
We use the theory of urns (a common framework in reinforcement learning) to
analyze existing methods for predictive policing. We show formally as well as
empirically why these methods will not work. Subsequently, we provide remedies
that can be used directly with these methods in a black-box fashion that improve
their behavior, and provide theoretical justification for these remedies.

\section{Related Work}
\label{sec:related-work}

Our work builds most strongly on the work of \citet{LumIsaac2016} described above, demonstrating the consequences of feedback loops in simulation in the predictive
policing setting.  There are a number of systems currently in place for predictive policing \cite{RAND}. The most well known system used for predictive policing is called \predpol \citep{predpol} (described in more depth below). Our implementation of \predpol is the one used by \cite{LumIsaac2016} in their work.  Recidivism prediction systems are also related to this work in that we believe they may exhibit some of the same feedback loop issues, given that recidivism outcomes are only known for prisoners who are released.  While the details of the actual implementations (such as COMPAS \cite{compas}) remain proprietary, \cite{recidivism} provide a
comprehensive review of the methods used in this area.



\subsection{\predpol}
The predictive policing software \predpol will be critical to our experimental investigations, so we describe it in more detail here.  \predpol \citep{predpol} assumes that crimes follow an earthquake
aftershock model, so that regions that previously experienced crime are likely
to experience crime again, with some decay.  \citet{predpol} model the crime
rate $\lambda_r(t)$ in region $r$ at time $t$ as follows:
$ \lambda_r(t) = \mu_r + \sum_{t_r^i < t} \theta \omega e^{-\omega(t -
    t_r^i)} $
 where $t_n^i$ represents the time of an event in region $r$, $\omega$ quantifies the time decay of a shock, and $\theta$
captures the degree to which aftershocks are generated from an initial
event. They use an expectation-maximization procedure to determine the
parameters of the model. 

Note that this model only uses incident data (including both discovered and
reported incidents -- see Section~\ref{sec:simple-pred-polic}) per region to determine the true crime rate\footnote{\predpol --- critically --- conflates amount of
crime and incident data.}
and does not use any context in the form of demographics, arrest profiles and so on. 
\predpol, in essence, is predicting where incidents will be reported or discovered (since that's all it sees), not where \emph{crime} will happen.
Each day officers are sent to the areas with
highest predicted intensity and the resulting discovered incident data is fed back into
the system.


\section{Predictive Policing with Urns}
\label{sec:predictive-policing}


We will model the predictive policing process by a series of urn models with increasing
complexity. Urn models (especially the \polya{}-Eggenberger urns) have a long
history in machine learning, but notably also 
in reinforcement learning \citep{pemantle2007}, where they have been used, starting
with the work of \citet{erev1998predicting}, as a way to model how bounded-rationality players in a game might
interact with each other. Studying the dynamics of urn models allows us to
understand the convergent behavior of reinforcement learning in such settings. 

We will use a \emph{generalized} \polya
urn model \citep{pemantle2007} containing balls of two colors (red and
black). At each time step, one ball is drawn from the urn, the color is noted, and
the ball is replaced. Then the following replacement matrix is used to decide
how to update the urn contents:
\[ \bordermatrix{ & \text{Red addition} & \text{Black addition} \cr 
\text{Sample red} & a & b \cr
\text{Sample black} & c & d } \]
This matrix says that if we initially sampled a red ball, then we
replace it and add $a$ more red balls and $b$ more black balls to the urn. 
We refer to the \emph{standard \polya urn} as a generalized urn with $a = d = 1$
and $b = c = 0$.

\subsection{Goals and assumptions}
\label{sec:simple-pred-polic}

In the simplest predictive policing setting, a precinct has a single police officer and polices two 
regions \one and \two. Every day the police officer is sent to one neighborhood
where they may or may not observe an incident; if they do, it is logged and we
refer to such a log as a \emph{discovered} incident. In addition, residents
might report incidents that are also logged: we call these \emph{reported}
incidents. 
The goal is to build a predictive model for where to send the officer on each day.  Specifically, the goal is to distribute the
police officers in proportion to the crime in each area.\footnote{Why should this be the goal?  Suppose there are exactly enough police officers to stop all the crime and no more, then a deployment according to the true crime rates will perfectly police all regions.}
\begin{goal}[Effective Policing]
A region with $\Lambda$ percent of the crime in the precinct should receive $\Lambda$ percent of the police.
\end{goal}
\noindent Achieving this goal requires learning the relative crime rates of the regions.

To understand the behavior of predictive models, we will make some simplifying
assumptions. We will firstly assume that the predictive model only uses current
statistics (in some form) to make predictions. 

\begin{assume}[Predictive Model]
  The officer tosses a coin based on current statistics to decide where to go
  next. 
\end{assume}

To fully specify a predictive model, we also need to understand \emph{context}
-- what information is collected during policing -- and \emph{ground truth} --
what assumptions we make about underlying crime rates.  We assume the simplest form of context.

\begin{assume}[Context]
  The only information retained about a crime is a count.\label{assume:context}
\end{assume}

Assumptions about ground truth are both critical and complicated.  For some
neighborhood \one, let $\lambda_\one$ be the underlying ground truth crime rate
for the neighborhood.  We will assume that this is observed via discovered and
reported incidents.  Let $d_\one$ be the rate at which police that visit
neighborhood \one discover incidents.  Let $w_d$ be the weight of the discovered
incidents within all incidents.  Similarly, let $r_\one$ be the rate at which
incidents are reported from neighborhood \one, and let $w_r$ be the weight of
reported incidents among all incidents.  We will assume that $w_r + w_d = 1$.
The total rate of incident data from neighborhood \one is then $w_d \cdot d_\one
+ w_r \cdot r_\one$. We note here that \emph{discovered} incidents are directly
implicated in the feedback loop since police are deployed in areas based on the
results of the predictive model. \emph{Reported} incidents on the other hand are
not.

To start our examinations, we make the following assumptions. In the subsections below, we explore what happens as we vary these factors. 

\begin{assume}[Truth in Crime Data]
\label{ass:discover_truth}
If an officer goes to a location \one with an underlying ground truth crime rate of $\lambda_\one$, the officer discovers crime at a rate of $\lambda_\one$.  I.e., $d_\one = \lambda_\one$.  Reported incidents are also reported at a rate that matches the underlying ground truth crime rate, i.e., $r_\one = \lambda_\one$.
\end{assume}

Note that Assumption \ref{ass:discover_truth} allows the predictive policing system to operate in a generous context.  There are many reasons to believe that this assumption does not hold.  We will show that even in this optimistic setting problems occur.

\begin{assume}[Discovery Only]
\label{ass:discovered}
Incident data is only collected by an officer's presence in a neighborhood.  Neighborhoods with no officers will contribute no incidents to the data.  I.e., $w_d = 1$ and $w_r = 0$.
\end{assume}

We will also start with the assumption that all incident data is made up of discovered incidents.  We will modify this assumption to also account for reported incidents in Section \ref{sec:reported}.

\subsection{Uniform crime rates}

Let us start by assuming that the crime rate is uniform between areas. 
\begin{assume}[Uniform Crime Rate]

\label{assume:uniform}
  If an officer goes to a location, crime happens with probability $\lambda$.  I.e., for any neighborhoods \one and \two, $\lambda_\one = \lambda_\two = \lambda$.
\end{assume}

Consider an urn that contains red and black balls, where the proportion of red
and black balls represent the current observed statistics of crime in areas
\one and \two respectively. Visiting area \one corresponds to picking a red ball
and visiting area \two corresponds to picking a black ball.  Observing crime
(which happens with probability $\lambda$)
causes a new ball of the same color to be placed in the urn. The initial balls are
always returned to the urn. 
The long-term distribution of red and black balls in the urn corresponds to the
long-term belief about crime prevalence in the two areas.

In general, we can describe the evolution of this process as the following urn. We toss a coin that returns $1$ with
probability $\lambda$. If the coin returns $1$, we simulate one time step of a standard \polya urn, and
if $0$, we merely replace the sampled ball. This corresponds to a standard \polya urn ``slowed'' down by a factor
$\lambda$. As such, its long-term convergence is well-characterized. Let the
beta distribution $\text{Beta}(\alpha, \beta)$ be a distribution over the
interval $[0,1]$ where the probability of $x$ is given by\footnote{The constant
  of proportionality is $\Gamma(\alpha)\Gamma(\beta)/\Gamma(\alpha + \beta)$
  where $\Gamma(x)$ is the standard gamma function.} $f(x; \alpha, \beta) \propto
x^{\alpha - 1} (1-x)^{\beta - 1}$

\begin{lemma}[\cite{renlund2010generalized}]
  Assume the urn starts off with $n_r$ red balls and $n_b$ black
  balls. Then the limiting probability of seeing a red ball is a draw from the beta
  distribution $\text{Beta}(n_r, n_b)$.
\end{lemma}

\paragraph{Significance.}
The long-term probability of seeing red is the long-term estimate of crime in
area \one \emph{generated by the model}. The above result shows that this probability is a random draw
governed only by the parameters $n_r, n_b$, which represents the prior
\emph{belief} of the system. In other words, the prior belief coupled with the
lack of feedback about the unobserved region 
\emph{prevents the system from learning that the two areas are in fact
identical with respect to crime rates. }

On the contrary, consider how this process would work \emph{without} feedback. The officer could
be sent to an area chosen uniformly at random each day, and this process would
clearly converge to a uniform crime rate for each area. Indeed, such a process
resembles the standard update for the bias of a coin where the prior
distribution on the bias is governed by a $\text{Beta}$ distribution.

\subsection{Non-uniform crime rates}
\label{sec:nonun-crime-rates}

Let us now drop the assumption of uniformity in crime rates, replacing Assumption~\ref{assume:uniform} by 

\begin{assume}[Non-uniform Crime Rate]
  A visit to area \one has probability $\lambda_{\one}$ of encountering a crime,
  and a visit to area \two has probability $\lambda_{\two}$ of encountering a
  crime. 
\end{assume}

Nonuniform crime rates in neighborhoods \one and \two can also be modeled by a \polya urn, with the caveat that the
updates to the urn are now random variables instead of deterministic
updates. Formally, we can think of the urn as being described by the $2\times 2$
(addition) matrix

\[
  \begin{pmatrix}
  X_\one & 0 \\
0 & X_\two  
  \end{pmatrix}
\]
where $X_\one$ is a  Bernoulli variable taking the value $1$ with probability
$\lambda_\one$ and $0$ with probability $1-\lambda_\one$, and $X_\two$ is
defined similarly

If the urn satisfied the so-called  \emph{balance}
condition that the number of balls added at each time step is a
constant \citep{mahmoud2012exactly}, then we could invoke standard results to
determine the limiting behavior. This
is not the case in this setting. However, we now show that it is possible to reduce this to
a deterministic update model by exploiting the Bernoulli structure of the
update. 

At any time $t$, let $n^{(t)}_\one, n^{(t)}_\two$ be the number of balls 
``colored'' \one and \two respectively. The probability of adding \emph{any} ball
to the urn is given by the expression 
\[ \frac{n^{(t)}_\one \lambda_\one + n^{(t)}_\two \lambda_\two}{n^{(t)}_\one +
    n^{(t)}_\two} \]
Note that this can be viewed as a convex combination of the two probabilities
$\lambda_\one$ and $\lambda_\two$ and so the overall probability of a ball being
added to the bin varies between two constants. 

As before, consider the update process limited to time steps when a ball is
added to the urn. The probability of adding a ball colored \one,
\emph{conditioned on adding some ball}, is given by 
\[ \frac{\Pr(\text{adding a \one{}-colored ball})}{\Pr(\text{adding some ball})}
  = \frac{n^{(t)}_\one \lambda_\one}{n^{(t)}_\one \lambda_\one + n^{(t)}_\two
    \lambda_\two}\]
with a similar expression for adding a \two{}-colored ball. 

This is identical to the deterministic \polya urn in which we sample an
$i$-colored ball, replace it and then \emph{add in $\lambda_i$ more balls of the same
color}. Essentially by conditioning on the event that we add a ball, we have
eliminated the randomness in the update while retaining the randomness in the
sampling. 

This latter \polya urn can be represented by the stochastic addition matrix

\begin{equation}
  \begin{pmatrix}
    \lambda_\one & 0 \\
    0 & \lambda_\two
  \end{pmatrix}\label{eq:second-urn}
\end{equation}

A very elegant result by \citet{renlund2010generalized} provides a general expression for the
long-term probability of seeing a \one{}-colored ball. 

\begin{lemma}[\cite{renlund2010generalized}]
\label{lemma:magic}
Suppose we are given a \polya urn with replacement matrix of the form 
\[
  \begin{pmatrix}
    a & b \\
    c & d
  \end{pmatrix}
\] with a positive number of balls of each kind to start with. Assume that
$a,b,c,d \ge 0$ and at least one entry is strictly positive. Then the limit of
the fraction of balls of each type exists almost surely. The fraction $p$ of
\one{}-colored balls can be characterized as follows:
\begin{itemize}
\item If $a = d, c = b = 0$, then $p$ tends towards a beta distribution. 
\item If not, then $p$ tends towards a single point distribution $x^*$, where
  $x^* \in [0,1]$ is a root of the quadratic polynomial
\[ (c+d-a-b)x^2 + (a - 2c-d)x + c. \] If two such roots exist, then it is the
one such that $f'(x^*) < 0$.
\end{itemize}
\end{lemma}

By limiting ourselves to the subsequence of events when some ball is added to
the urn, and using the above general lemma characterizing the asymptotics of
\emph{deterministic} urn updates from \citet{renlund2010generalized}, we have the following lemma about the urn under this new assumption. 

\begin{lemma}
\label{lem:urn_skew}
In the urn with addition matrix given above, the asymptotic
probability of sampling a red ball is $1$ if $\lambda_\one > \lambda_\two$ and
is zero if $\lambda_\two > \lambda_\one$. 
\end{lemma}
\begin{proof}
  Invoking Lemma~\ref{lemma:magic}, we set the parameters $b = c = 0$, $a =
  \lambda_\one$ and $d = \lambda_\two$. The resulting quadratic polynomial is
  $(\lambda_\two-\lambda_\one)x^2 + (\lambda_\one - \lambda_\two)x = 0$. This
  polynomial has two roots: $x = 0, 1$. The first derivative is $(\lambda_\two -
  \lambda_\one)(2x-1)$. If $\lambda_\one > \lambda_\two$, then this is negative
  for $x = 1$. Conversely, if $\lambda_\two > \lambda_\one$, then this is
  negative for $x= 0$.
\end{proof}

\paragraph{Significance.}
In this scenario, the update process will view one region as having much more
crime than the other, \emph{even if crime rates are similar}. In particular,
if region \one has a crime rate of $10\%$ and region \two  has a crime rate of
$11\%$, the update process will settle on region \two with probability $1$. 
This is a classic ``go with the winner'' problem where feedback causes a
runaway effect on the estimated probabilities. 

\subsection{Accounting for reported incidents}
\label{sec:reported}

Now we consider what happens when we remove Assumption \ref{ass:discovered},
i.e., we allow both discovered and reported incidents to be used as input to the
urn model as is more usually the case in predictive policing systems. As
discussed earlier, the weight terms $w_d$ and $w_r$ are used to weight
discovered and reported crimes from a neighborhood, and so the total weight of
crime incidents from (say) area \one would be $w_d d_\one + w_r r_\one$ if it
was visited, and $w_r r_\one$ otherwise. This leads to the following urn
replacement matrix:

\[
  \begin{pmatrix}
  w_d d_\one + w_r r_\one & w_r r_\two \\
  w_r r_\one & w_d d_\two + w_r r_\two  
  \end{pmatrix}
\]

\noindent where, as in Section~\ref{sec:nonun-crime-rates}, we should interpret the entries
of the matrix as expected values of a Bernoulli process.

Using the same trick as
in Section~\ref{sec:nonun-crime-rates}, we can reinterpret the above matrix as a \emph{deterministic} update
process, and invoke 
Lemma~\ref{lemma:magic} to understand the limiting behavior. The corresponding
quadratic equation associated with this replacement urn is given by:
\begin{multline*}
   f(x) =  w_d(d_\two - d_\one)x^2 + (w_d(d_\one - d_\two) - \\ w_r (r_\one +
   r_\two))x  +   w_r r_\one = 0
\end{multline*}
\noindent Let $R = w_r (r_\one + r_\two)$ be the total weight of reported incidents, and let us
denote $w_d(d_\two - d_\one)$ by $\Delta_d$, the weighted \emph{differential} in
discovered crime. We can then rewrite the above expression as:
\[ f(x) = \Delta_d x^2 - (\Delta_d + R)x + w_r r_\one = 0\]

We can now find the roots of $f(x)=0$. These are given by 

\[ x = \frac{(\Delta_d + R) \pm \sqrt{(\Delta_d + R)^2 - 4 \Delta_d w_r r_\one}}{2\Delta_d} \]
which can be written as 
\[ x = \nu \pm \sqrt{\nu^2  - \frac{w_rr_\one}{\Delta_d}} \]
where $\nu = \frac{1}{2} + \frac{R}{2\Delta_d}$. 
Taking the first derivative,
\[ f'(x) = 2\Delta_d x - (\Delta_d + R) \]
and thus $f'(x) < 0$ when $x < \frac{1}{2} + \frac{R}{2\Delta_d}$. Therefore, by
Lemma~\ref{lemma:magic}, the limiting fraction of ``$\one$-colored'' balls in
the urn is 
\begin{equation}
 x^* = \nu - \sqrt{\nu^2  - \frac{w_r r_\one}{w_d(d_\two - d_\one)}}\label{eq:1}
\end{equation}

\subsubsection{Interpretation}
\label{sec:interpretation}

We can interpret Equation~(\ref{eq:1}) through a number of cases. Firstly,
consider the case of \emph{no feedback}. This corresponds to setting $w_d =
0$. In that case, the urn replacement matrix is fixed: regardless of which ball
we draw, we always add $r_\one$ $\one$-colored and $r_\two$ \two{}-colored
balls. Clearly, the limiting fraction of $\one$-colored balls is $\frac{r_a}{r_a
  + r_b}$ and this is the answer we would expect given the crime reporting rates -- we denote this fraction as
$\lambda^*$. 


We can rewrite Equation~(\ref{eq:1}) in terms of $\lambda^*$ by introducing a
change of variable. Define $\kappa = R/\Delta_d$ which allows us to rewrite $\nu =
(1 + \kappa)/2$. We can now rewrite Equation~(\ref{eq:1}) as

\begin{equation}
 x^* = \frac{1+\kappa}{2} - \sqrt{\paren[\Big]{\frac{1+\kappa}{2}}^2  -
   \lambda^* \kappa}.\label{eq:2}
\end{equation}
The second term under the square root comes from noting that $w_r
r_\one/\Delta_d = r_\one/(r_\one + r_\two) \cdot w_r (r_\one +
r_\two)/\Delta_d$. 

A first observation is that as $\lambda^* \rightarrow 0$, $x^* \rightarrow
\lambda^*$. Similarly, as $\lambda^* \rightarrow 1$, $x^* \rightarrow
\lambda^*$. In other words, if the crime rates between the neighborhoods are
heavily skewed, this urn will converge to a good approximation of the correct
answer. 

For intermediate values of $\lambda^*$, we transform the equation as follows:
\begin{align*}
   x^* &= \frac{1+\kappa}{2} - \sqrt{\paren[\Big]{\frac{1+\kappa}{2}}^2  -
   \lambda^* \kappa} \\
  &= \frac{1+\kappa}{2}\paren[\Big]{1 - \sqrt{1  -
   \frac{\lambda^* \kappa}{\paren[\Big]{\frac{1+\kappa}{2}}^2}} }
\intertext{which in  the limit, as $\kappa$ grows, can be expressed as}
   x^* &= \frac{1+\kappa}{2}\paren[\Big]{\frac{\lambda^*
      \kappa}{2\paren[\Big]{\frac{1+\kappa}{2}}^2}}\\
\intertext{by a binomial approximation, yielding}
  x^* &= \lambda^* \frac{\kappa}{\kappa+1} = \lambda^* \frac{R}{R + \Delta_d}
\end{align*}

\paragraph{Significance}

The limiting behavior of this urn is represented by $x^*$. How does this relate
to the ideal limiting behavior $\lambda^*$? For $x^* \approx \lambda^*$, it must
be that the ratio $R/(R + \lambda_d)$ is close to $1$. This can happen in two
ways. Either $R$ must be very large, or $\Delta_d$ must be
small. $R = w_r(r_a + r_b)$ which is bounded by $2$. Thus, the only other option
is to have $\Delta_d$ be very small. Recall that $\Delta_d = w_d(d_{\two} -
d_{\one})$. To make it small, we must either make $w_d$ small, which corresponds
to discounting the importance of discovered incidents (thus relying heavily on the distribution of 
reported incidents assumed to be correct by Assumption \ref{ass:discover_truth}), or it must be that the
discovered crime rates $d_\two$ and $d_\one$ are very similar. In other words,
the only scenarios where feedback does \emph{not} drive the outcome away from
the true result are when we effective ignore feedback (by downweighting the
importance of discovered crime) or when the crime rates are similar enough for
the feedback to not matter. However, it is precisely when crime rates are
different that predictive policing is of value (because resources are then
deployed differently). Thus, once again the urn model reveals problems (via
simulation) in existing models for predictive policing.

\subsection{Modifying the urn model to account for feedback}
\label{sec:modifying-urn-model}

\looseness-1 In order to learn the crime rate, we want the \polya urn to contain balls in proportion to the relative
probability of crime occurrence. As we have seen, a standard
P\'{o}lya urn with stochastic update rates will converge to a distribution that
has no relation to the true crime 
rates. Here, we present a simple change to the urn process
which \emph{does} guarantee that the urn proportion will converge to
the ratio of replacement (i.e. crime) rates.

\subsubsection{Discovered Incidents Only}

Again, we first consider what happens if Assumption \ref{ass:discovered} is in place.

Consider the standard \polya urn update rule: the probabilities
$\lambda_\one$ and $\lambda_\two$ model the probability of an additional
ball being added to the urn, \emph{conditional} on a ball of the
respective color having been sampled. This means that the
probability of a ball being added is not $\lambda_\one$, but
$\lambda_\one \frac{n^{(t)}_\one}{n^{(t)}_\one + n^{(t)}_\two}$. As a result,
the expected fraction of $\one$-balls being added to the urn after one step of the process
is $\frac{\lambda_\one n_\one}{\lambda_\one n_\one + \lambda_\two n_\two}$
  instead of $\frac{\lambda_\one}{\lambda_\one + \lambda_\two}$. 

This immediately suggests a fix: instead of always adding the new balls, we \emph{first sample another ball from the urn, and only add
  the new balls if the colors are different}.
With this fix, the probability of adding a ball with label
\one is $\frac{n^{(t)}_\one}{n^{(t)}_\one + n^{(t)}_\two}\lambda_\one\frac{n^{(t)}_\two}{n^{(t)}_\one + n^{(t)}_\two}$, while the
probability of adding a ball with label \two is
$\frac{n^{(t)}_\two}{n^{(t)}_\one + n^{(t)}_\two}\lambda_\two\frac{n^{(t)}_\one}{n^{(t)}_\one + n^{(t)}_\two}$. Crucially, these two
expressions are proportional to $\lambda_\one$ and $\lambda_\two$, except
for a constant factor that is a function of the current state of the
urn.

The intuition behind this fix is that if our decision procedure sends
police to region \one 90\% of the time, we should not be surprised
that discovered incidents in region \one happen at a rate of nine to
one, even if the crime rate is the same across both regions. In such a
scenario, if we see a crime in region \one (where police go 90\% of the
time), we should simply drop the incident record 90\% of the time;
analogously, in region \two (where police only go 10\% of the time),
we drop the incident record 10\% of the time.

One way to interpret our fix is as a form of \emph{rejection
  sampling}: we are dropping sampled values according to some
probability scheme to affect the statistic we are collecting.  The
importance-sampling analog to this scheme would be to use
\emph{weighted balls}, where the weight of each ball is inversely
proportional to be rate at which police are sent. Effectively, we want
a scheme where as more police are sent, smaller weights are assigned
to discovered incidents. But such a scheme is precisely the
Thompson-Horvitz estimator, used in survey designs with unequal
probability distributions~\citep{horvitz1952generalization}, and so we
see that our proposal is a rejection-sampling variant of
Thompson-Horvitz estimation.

\subsubsection{Reported and Discovered Incidents}
\label{sec:rep-and-disc}
Now we consider what happens if there are both discovered and reported incidents.  Intuitively, we want to correct for the runaway feedback caused by the discovered incidents, but not over-correct for the reported incidents, which don't suffer from the issue.  Recall that the replacement matrix is:
\[
  \begin{pmatrix}
  w_d d_\one + w_r r_\one & w_r r_\two \\
  w_r r_\one & w_d d_\two + w_r r_\two  
  \end{pmatrix}
\]
Suppose that Assumption \ref{ass:discover_truth} is in place and recall that $w_d + w_r = 1$.  Then this replacement matrix is:
\[
  \begin{pmatrix}
  \lambda_\one & w_r \lambda_\two \\
  w_r \lambda_\one & \lambda_\two  
  \end{pmatrix}
  = \]
  \[
    \begin{pmatrix}
  w_d d_\one & 0 \\
  0 & w_d d_\two
  \end{pmatrix}
  +
      \begin{pmatrix}
  w_r r_\one & w_r r_\two \\
  w_r r_\one & w_r r_\two  
  \end{pmatrix}
\]
Note that the first matrix represents the discovered replacement and the second
represents the reported replacement.  
From the previous section, we know how to modify the discovery replacement
matrix so that the feedback effect is mitigated.  We first apply that same
technique here, \emph{but only to the discovered incidents}. As before, doing
this ensures that the replacement contributes (in expectation) exactly $w_dd_\one$
to the urn when visiting $\one$, and $w_dd_\two$ when visiting $\two$. 

But what about the reported incidents? If we add them as is (i.e as given by the
second matrix), the total contribution in the case of an $\one$-visit is
$w_dd_\one + w_r r_\one + w_r r_\two$. Again, invoking
Assumption~\ref{ass:discover_truth}, the total contribution becomes
$\lambda_\one + w_r \lambda_\two$ (with a similar expression for a
$\two$-visit). Unfortunately, this expression leads to the urn converging
to an incorrect rate, ultimately because the contribution to the region not visited has
been incorrectly down-weighted. The fix is simple: we remove the
down-weighting of the reported incidents in the neighborhood where police were
not deployed.  The resulting replacement matrix, in expectation, is:
\[
  \begin{pmatrix}
  w_d d_\one + w_r r_\one & r_\two \\
  r_\one & w_d d_\two + w_r r_\two  
  \end{pmatrix}
\]
and we apply our earlier fix to any \emph{discovered} data.  This ensures that in expectation, the contribution to the urn
\emph{regardless} of whether $\one$ or $\two$ is visited
 is $\lambda_\one$ $\one$-balls and
$\lambda_\two$ $\two$-balls, as desired. 



\section{Evaluating the urn model}

\begin{figure*}[t]
  \centering  
\begin{tabular}{ccccc}
& \multicolumn{2}{c}{Without Improvement Policy} & \multicolumn{2}{c}{With Improvement Policy} \\
& Top1 vs. Top2 & Top1 vs. Random & Top1 vs. Top2 & Top1 vs. Random \\
\rotatebox{90}{~~~Discovered Only} &
\includegraphics[width=1.3in]{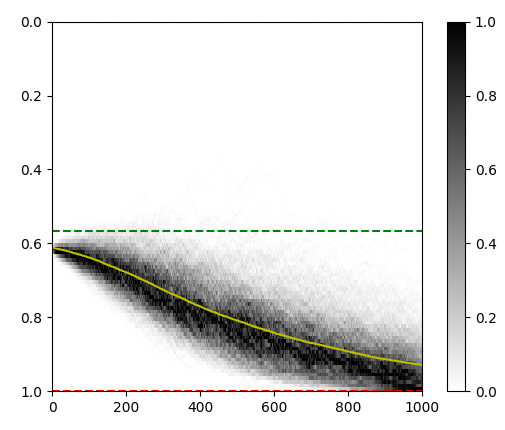} &
\includegraphics[width=1.3in]{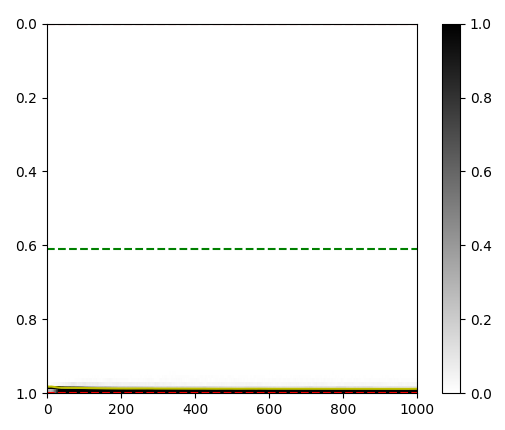} &
\includegraphics[width=1.3in]{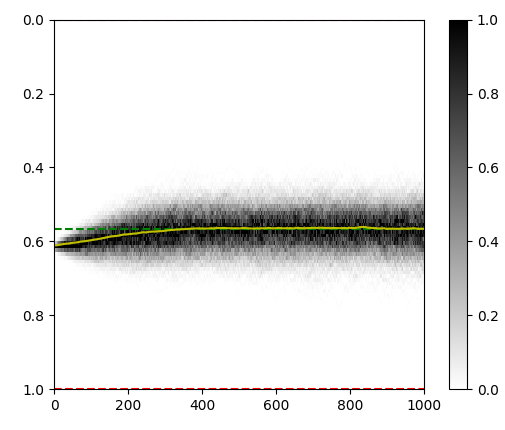} &
\includegraphics[width=1.3in]{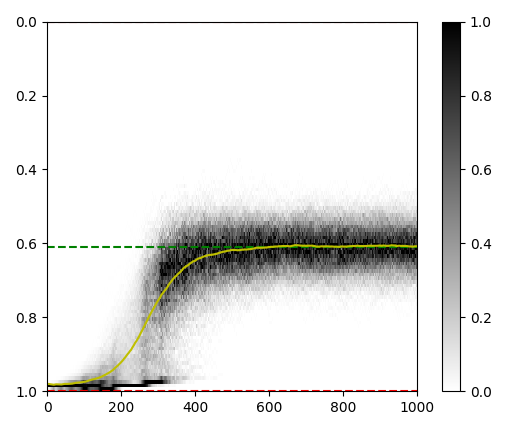}
\\
\rotatebox{90}{~~~~All Incidents} &
\includegraphics[width=1.3in]{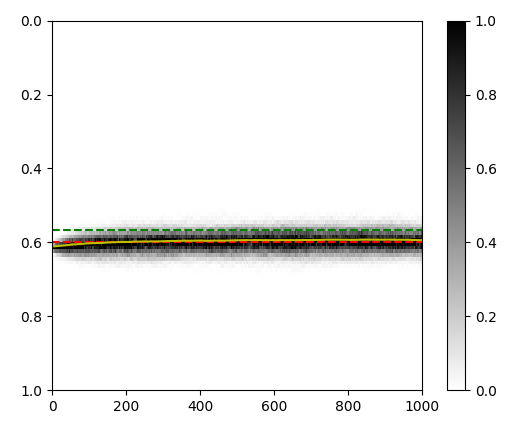} &
\includegraphics[width=1.3in]{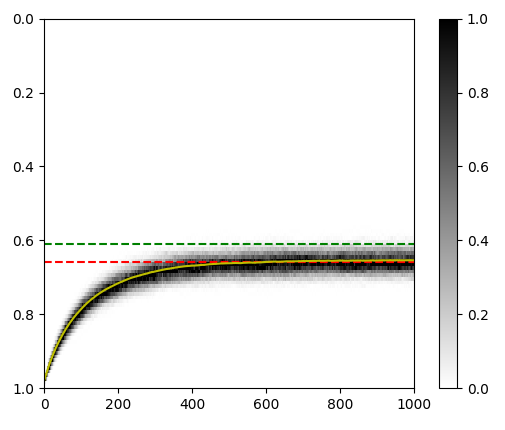} &
\includegraphics[width=1.3in]{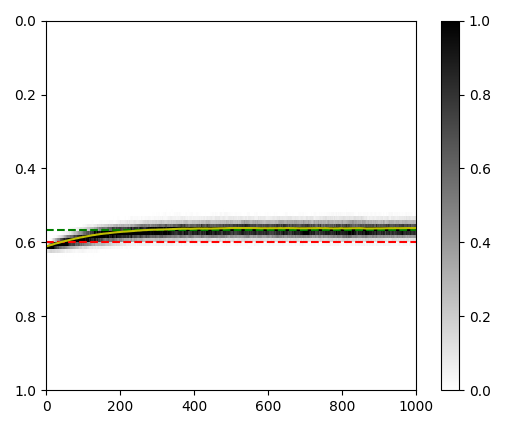} &
\includegraphics[width=1.3in]{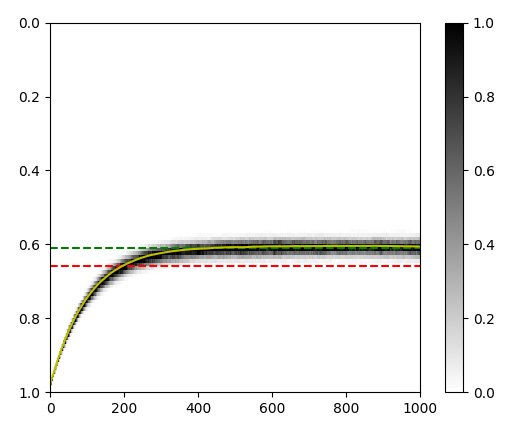}\\
\end{tabular}
  \caption{The distribution over 1000 days versus percentage of balls from region $Top1$ in the urn over 1000 runs.  A police force deployed based on the underlying crime rates would send $56.7\%$ of the force to $Top1$ instead of $Top2$ and $61.0\%$ of the force to $Top1$ instead of $Random$ (the green line shown).  Top row (discovered incidents only): both charts (left) converge to sending $100\%$ of the force to $Top1$, while with the improvement policy (right) the charts appear to converge to the correct crime rates.  Bottom row (all incidents, equally weighted): both charts (left) converge to the incorrect rate (red line), while with the improvement policy (right) the charts appear to converge correctly to the true crime rates.}
  \label{fig:polya_experiments}
\end{figure*}

In this section, we will focus on validating the existence of the feedback loop problem experimentally within our urn model. Code for our urn simulations can be found at \url{https://github.com/algofairness/runaway-feedback-loops-src}.

\subsection{Observational decay}
\label{sec:initial-bias-can}

Thus far, our urn models have captured some key elements of the model used by \predpol -- 
the idea of differential crime rates as well as the updates based on discovered and reported incidents. 
\predpol also includes a notion of \emph{limited memory}, both by
incorporating time decay into crime aftershocks, and by using a limited time
window for training. We model limited memory in the urn setting by adding a
simple notion of decay. After every round, each ball disappears from the urn
independently with a fixed probability $p_d$. This can be thought of as a
relaxation of Assumption~\ref{assume:context}. 
Varying $p_d$ is analogous to varying the size of the
\predpol training window. 

\subsection{Illustrating runaway feedback in urns}

To the best of our knowledge there is no
theoretical characterization of the asymptotic distributions in this full model once 
the notion of decay is included. We
 present empirical evidence illustrating the problems with using this model
to learn crime rates. In our experiments, $p_d = 0.01$. 

\looseness-1 Using the \citet{LumIsaac2016} data, we consider a two neighborhood police deployment scenario using, first, the two regions of Oakland with the most historical incident data ($Top1$ and $Top2$) and, second, the Oakland neighborhood with the most incidents as compared to a randomly chosen region with many fewer incidents ($Random$).  We simulate the effect of the historical incident data on the prior for the system by determining the number of balls for each region in our urn based on the past number of incidents.  We use the full number of incidents (609, 379, and 7 for regions $Top1$, $Top2$ and $Random$ respectively) as the starting number of balls in the urn from each region. 
The urns are then updated based on the estimated number of daily drug use incidents, i.e., $\lambda_{Top1} = 3.69$, $\lambda_{Top2} = 2.82$, and $\lambda_{Random} = 2.36$.

\subsubsection{Discovered Incidents Only}

First, we begin with Assumption \ref{ass:discovered} in place stating that we'll only consider discovered incident data (i.e., $w_r = 0$).  This allows us to isolate the part of the data that causes the feedback loop in order to examine its effect.

\looseness-1 The results, shown in the top left of Figure \ref{fig:polya_experiments},
demonstrate that even if police sent to a neighborhood discover crime incidents according
to the true crime rate (Assumption \ref{ass:discover_truth}), the urn model will converge to \emph{only} sending police to the neighborhood with the most crime.  This replicates (within our urn
model) the feedback loop problems with \textsc{PredPol} found by \citet{LumIsaac2016}.
Recall, from Lemma \ref{lem:urn_skew}, that skew occurs even if the difference in crime rates
between the two neighborhoods is small.  Note that  while we included a notion
of decay in our urn model in order to more closely model \textsc{PredPol}, we
found similar results under the urn model without decay. 

\subsubsection{Discovered and Reported Incidents}

Now, considering both reported and discovered incident data, we repeat the experiments.  Again, we'll assume that both discovered and reported incidents are reported at the underlying true rate (Assumption \ref{ass:discover_truth}), and we'll assume that these incidents are equally weighted, i.e., that $w_d = w_r = 0.5$.  The results shown in the bottom left of Figure \ref{fig:polya_experiments} show that while the error in police deployment is not as great as if only discovered incidents are used, the urn still does not converge to the correct rate.  Here, it's important to note the strength of the assumption that incidents are reported at the true underlying rate and not influenced by police deployment - we suspect that this assumption helps this convergence to be closer to (though still not the same as) the correct rate.

\subsection{Evaluating the modified urn}

Using this improvement policy to determine when to replace balls, we can now determine if the urns can learn the true crime rate despite the issue of observational bias.  Again, using the estimated daily drug usage per region as the underlying true crime rate and the historical incident data as the prior for the urn color distribution, we simulate the urn's ability to find the relative crime rates in two regions, the $Top1$ and $Top2$ incident regions and a $Random$ region.  As shown in the right of Figure \ref{fig:polya_experiments}, urns under this improvement policy converge to a distribution of colors that represents the true crime rate, whether using only discovered incidents or both discovered and reported incidents.

\begin{figure*}[t]
  \centering
  \includegraphics[width=5in]{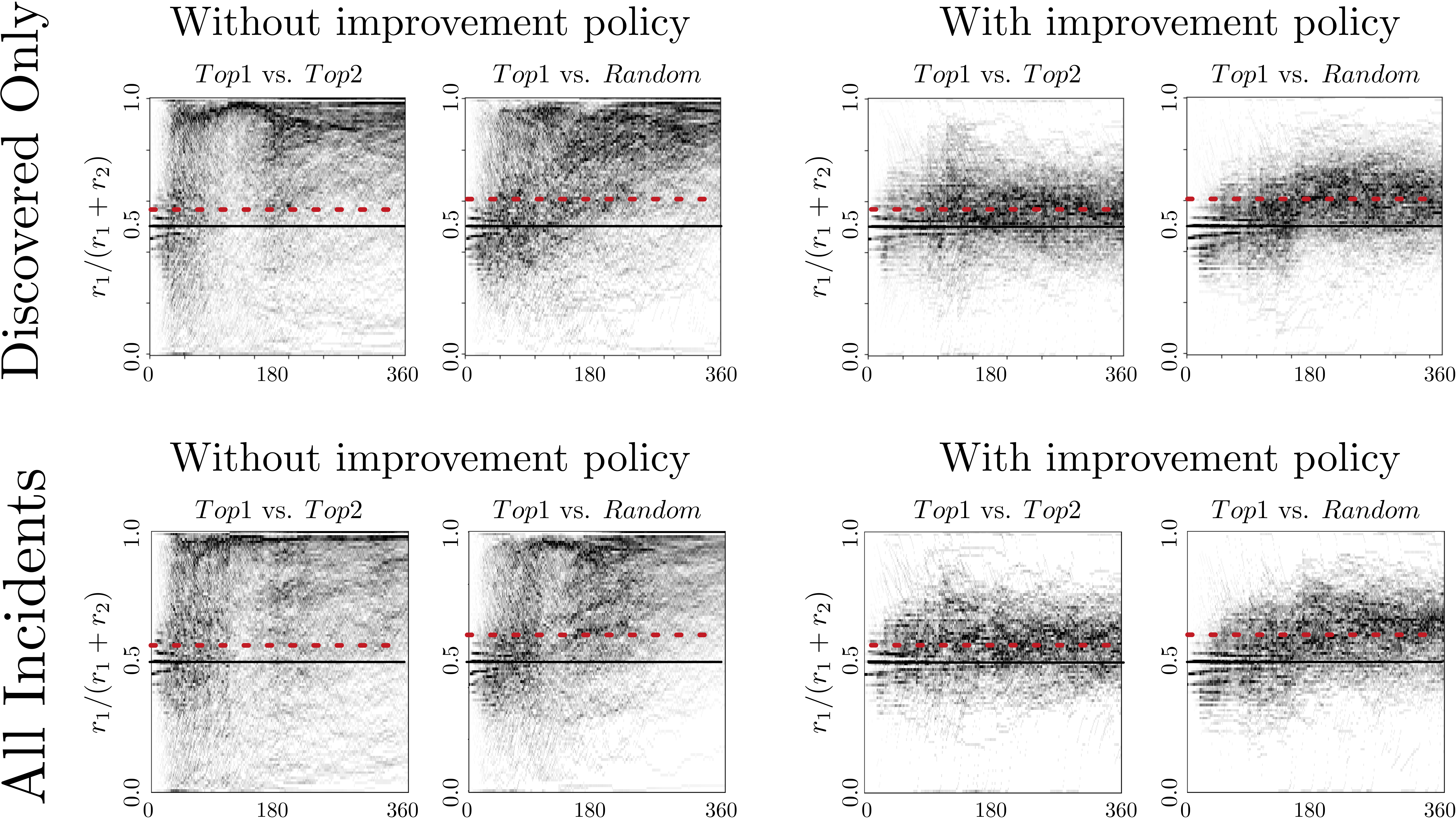}
  \caption{\textsc{PredPol}'s relative deployment to region $Top1$ versus $Top2$ or $Random$. Along the top row, we use the model which only accounts for discovered incidents (those based on police having been deployed to an area). Along the bottom row, we use the model which accounts for both discovered and reported incidents.
Left: \textsc{PredPol} operating as usual.  Right: discovered incident entries modified using our improvement policy.  Police deployment based on underlying crime rates would send $56.7\%$ of the force to $Top1$ instead of $Top2$ and $61.0\%$ of the force to $Top1$ instead of $Random$.  These correct crime rates (indicated with a dashed red line) appear to be what \textsc{PredPol} converges to under the improvement policy.}
\label{fig:predpol}
\end{figure*}


\section{Fixing \textsc{PredPol}}

\subsection{Modifying \textsc{PredPol} in a black-box manner}
The urn models we explore provide a justification for the observed feedback loop
failures of \textsc{PredPol}. But can we remedy \textsc{PredPol} itself using
the improvements described in Section~\ref{sec:modifying-urn-model}? We first
demonstrate how asymmetric feedback affects \textsc{PredPol} by simulating the
decisions a precinct might take after running it. 
We run \textsc{PredPol}'s prediction model (using the \citet{LumIsaac2016} data and implementation), trained on Oakland historical crime data, and generate crime according to the drug usage rates described above. 

At each simulation day, \textsc{PredPol} trains on the previous 180 days of incident data,
and produces predicted crime rates $r_\one$ and $r_\two$. The decision of where to send police is made probabilistically, by a
Bernoulli trial with $p=\myfrac{r_\one}{r_\one + r_\two}$. This models the targeting effect of sending
more police where more crime is expected, simulating a typical use of \textsc{PredPol}~\citep{predpol}.

To counteract the effects of the feedback, we can employ the same strategy as in
Section~\ref{sec:modifying-urn-model}. The key insight is that we need only
filter the inputs presented to \textsc{PredPol} rather than trying to modify its
internal workings. Specifically, once we obtain crime report data from the
system, we conduct another Bernoulli trial with
$p=\myfrac{r_O}{r_\one + r_\two}$, where $r_O$ is the predicted rate of the
district we did \emph{not} police that day, and \emph{only add the incidents to
  the training set if the trial succeeds}. In other words, the more likely it is
that police are sent to a given district, the less likely it is that we should
incorporate those discovered incidents.

\subsection{Evaluating the \textsc{PredPol} simulation and its repair}
Simulating the effects of \textsc{PredPol} on policing as described above, both
before and after our improvement policy is applied, we compare the policing
rates of region $Top1$ to $Top2$ and $Top1$ to $Random$ as before. 
Each simulation is repeated 300 times and run for one
year. As can be seen in the top row of Figure~\ref{fig:predpol}, regular \textsc{PredPol} rates fluctuate wildly
over different runs, and do not converge to the appropriate crime
rates (marked with the red dashed line). However, when the inputs to
\textsc{PredPol} are changed according to our improvement policy,
\textsc{PredPol}'s prediction rates appear to fluctuate around the correct crime
ratio. Note that the process is still quite noisy, a further indication that
\textsc{PredPol} generates crime rate predictions that are still somewhat
unreliable. 

In Section~\ref{sec:rep-and-disc}, we provided an analysis and correction for urn models based on more than only discovered incidents. We provide a similar analysis for the mixed case in \textsc{PredPol}, shown along the bottom row of Figure~\ref{fig:predpol}. Note that even with a large number of reported incidents, \textsc{PredPol} seems to remain susceptible to runaway feedback. When the correction mechanism of Section~\ref{sec:rep-and-disc} is applied to the (discovered only) incidents, \textsc{PredPol} appears to converge to the appropriate crime rate predictions.

\section{Discussion and Limitations}
In this paper we show that urn models can be used to formally model predictive
policing as well as indicate remedies for problems with feedback. We demonstrate
this both formally and empirically. Our solution also suggests a black-box
method to counteract runaway feedback in predictive policing by appropriately
filtering the inputs fed to the system.  

Our results are not just a qualitative indicator of problems with feedback. They
also indicate exactly how the problem of runaway feedback can be exacerbated:
specifically as crime rates vary between regions and as the model relies more
and more on discovered incident reports. Our results also indicate that if crime
rates are more or less the same between regions, then the problem of feedback is
much less, and it might be possible to generate reasonable predictions without
explicitly countering feedback loops (though the results will still be inaccurate). 

There are many avenues that our analysis does not yet explore. Firstly, while we
expect that our solution  generalizes to multiple regions (and indeed the
problems with feedback remain exactly the same), there might be technical
difficulties in working with the much smaller probabilities we will
encounter. As an abstraction of predictive policing, an urn model suffices to
capture feedback issues, but does not account for potentially richer predictive
systems that might use other information (for example demographics) to make
predictions. Another interesting dimension that is unexplored is the fact that
different types of crime might have different reporting and discovery profiles,
and might interact with each other in a predictive model in complex ways. 

One of the most crucial assumptions we make (and one that in fact is sympathetic
to current predictive policing models) is that reported and discovered incident rates
track the true crime rates. There is considerable evidence that crime reporting
is noisy and  skewed by area and by type of crime \citep{fte}. Once we remove that
assumption, the analysis becomes more complicated, and while the problems of
runaway feedback remain, the solutions might not continue to work. In this case,
we would require better models to describe how crime rates manifest themselves
in terms of reported and discovered incidents.


\section{Acknowledgements}
This paper would not have been possible without Kristian Lum and William Isaac's generosity in sharing the code and data developed for \citep{LumIsaac2016}.  Many thanks!

\bibliography{refs}


\end{document}